# Rare-earth doped optical dimers

**Galina Nemova**

*Department of Engineering Physics, Polytechnique Montréal, PO Box 6079, Station Centre-Ville, Montréal, Canada, H3C 3A7*
*galina.nemova@videotron.ca*

**Abstract:** The optical forces applied to a rare-earth doped optical dimer made of identical $Yb^{3+}$:YAG nanospheres placed in the plane electromagnetic field propagating in vacuum has been theoretically considered. The electromagnetic fields at wavelengths $968\ nm$ and $1030\ nm$ have been normally and axially directed to the dimer axis. Wavelengths $968\ nm$ and $1030\ nm$ are resonant with the electron transitions of the $Yb^{3+}$ ions and can generate the Stokes and anti-Stokes cycles, respectively. It has been shown that the electromagnetic field propagating at wavelength $1030\ nm$ not only can be used to control optical forces applied to the dimer, but can cause cooling of the dimer, which is desirable in a number of applications.

## 1. Introduction

The optical forces were recognized as early as 1619 by Kepler, when he first explained the deflection of comet tails by the rays of the sun [1]. After the invention of the laser, Arthur Ashkin observed that a micrometer particle in water could be accelerated by a laser beam, or trapped by two counterpropagating laser beams [2]. These experiments unveiled the laser beams as a tool for the optical manipulation of microsized objects with laser light. Now, optical manipulation with laser light is widely used in various disciplines including biophysics [3], atom physics [4], quantum science technology [5], and nanotechnology [6]. It is common knowledge that optical trapping is based on the interplay between the gradient and scattering forces [7]. Laser light illuminating a dielectric sub-wavelength particle polarizes it in phase with incoming electromagnetic wave. The induced dipole interacts with the field gradient and positions itself at the point of the highest field to minimize its interaction energy. This interaction is a source of the gradient force. Scattering force arises due to the light momentum change upon the scattering from a particle. It generally points in the direction of the light propagation. Although the possibility for the manipulation of many different single samples using the laser light is well investigated [7-14], simultaneous light manipulation of more than one particle remains unexplored and interesting area of research for the creation of a self-organized structure of particles, which is a prime system for investigating complex non-equilibrium phenomena [15-18]. Nanoparticles with low-refractive-index, such as silica, polystyrene, have low permittivity. Low permittivity of nanoparticles is an obstacle on the way of development technologies that rely on refractive index mismatching between nanoparticles and the surrounding medium [19-21]. Metallic nanoparticles have high permittivity, but the heat generated from the metal increases the thermal fluctuations destabilizing the trap [22-24]. Semiconductor nanostructures with comparable permittivity demonstrate limited trapping force stiffness [25]. Nanocrystals doped with rare-earth (RE) ions were proposed as a solution to the problem however for some incident wavelengths they undergo heating as a result of Stokes effect [26].

The aim of this work is to investigate the optical forces applied to a dimer made of two identical low phonon RE-doped nanospheres placed into the propagating plane linearly polarized electromagnetic wave. Two schemes have been considered. In one of the schemes the electromagnetic wave propagates normally to the dimer axis; in the second one it propagates along the dimer axis (Fig. 1). Without loss of generality, a dimer made of $Yb^{3+}$:YAG has been considered. The forces applied to the $Yb^{3+}$:YAG dimer have been compared with the forces applied to the identical undoped YAG dimer. These forces have been estimated at wavelengths $968\ nm$ and $1030\ nm$. The electromagnetic field at wavelength $968\ nm$ is a source of the Stokes cycle in $Yb^{3+}$:YAG accompanied by heating of the sample. The field at wavelength $1030\ nm$ is a source of the anti-Stokes cycle accompanied by cooling.

The theoretical description of the process is presented in Section 2. The results of the simulations are discussed in details in Section 3.

## 2. Theoretical analysis

Consider a dimer composed of two identical optically bound isotropic dielectric spherical nanoparticles separated by a distance $R$ (Fig. 1). One of the spheres is placed at the center of the coordinate system. Consider two schemes. In one of the schemes the dimer is located along y-axis perpendicularly to the electromagnetic wave propagating along the z-axis (Fig.1a). In the second scheme the dimer is located in parallel to the propagating electromagnetic field (Fig.1b). The electromagnetic wave is a liner polarized plane wave with the electric field vector, $E_0$, parallel to the y-axis.

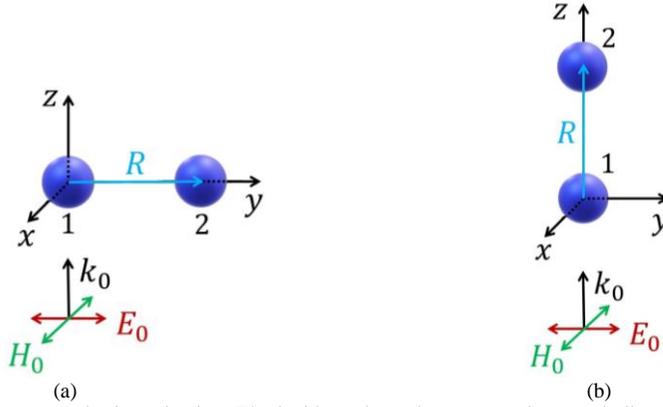

(a) (b)

Fig. 1. Structures under investigation. The incident plane electromagnetic wave is linearly polarized. It propagates normally to the dimer axis (a) and along the dimer axis (b). $E_0$ and $H_0$ are the amplitudes of the electric and magnetic fields, respectively.

Following Ref. [27] I consider the classical model of induced oscillating dipoles to express the forces acting upon nanospheres of the dimer. It is well known that the total electric field at the position of each dipole induced in a nanosphere is the sum of the incident field and the field emitted by the other dipole induced in the second nanosphere of the dimer. These total electric fields at the positions nanosphere1 and nanosphere 2 can be presented, respectively, as

$$E_i(0) = E_i^{in}(\vec{R}) + \alpha G_{ij}(0,\vec{R})E_j(\vec{R}), \tag{1a}$$

$$E_i(\vec{R}) = E_i^{in}(\vec{R}) + \alpha G_{ij}(\vec{R},0)E_j(0), \tag{1b}$$

where

$$G_{ij}(0,\vec{R}) = k_0^3 e^{ik_0 t}\left[\left(-\frac{1}{k_0 R} - \frac{i3}{k_0^2 R^2} + \frac{3}{k_0^3 R^3}\right)\frac{R_i R_j}{R^2} + \left(\frac{1}{k_0 R} + \frac{i}{k_0^2 R^2} - \frac{1}{k_0^3 R^3}\right)\delta_{ij}\right], \tag{1c}$$

is the electric dyadic Green's function, $E_i^{in}(0)$ and $E_i^{in}(\vec{R})$ are $i^{th}$ - component of the incident electric field at the origin of coordinates and at the point with the radius vector $\vec{R}$, respectively. Here $i$ and $j$ denote the Cartesian axes $x, y, z$. $k_0 = 2\pi/\lambda_p$, is the wavevector, $\lambda_p$ is the wavelength of the incident electromagnetic wave propagating in vacuum. The polarizability of the nanospheres of the dimer with the radius $a$ and the electric permittivity $\varepsilon_p$ has the form [28]

$$\alpha = \frac{\alpha_0}{1-i\frac{2}{3}\alpha_0 k_0^3}, \text{ where } \alpha_0 = a^3 \frac{\varepsilon_p - 1}{\varepsilon_p + 2}. \tag{2}$$

With a knowledge the total electric filed, $E_i(\vec{R})$, at the location of particle 2, the time-averaged $\gamma^{th}$- component of the optical force, $F_\gamma^{(2)}$, acting on the particle 2 can be expressed as

$$F_\gamma^{(2)} = \frac{1}{2} Re\left[\alpha E_i^*(\vec{r}_2) \frac{\partial E_i(\vec{r})}{\partial r_\gamma}\bigg|_{\vec{r}=\vec{r}_2}\right]. \tag{3}$$

Here $Re[Z]$ denotes the real part of $Z$. Substituting relation (1) in to relation (3) one can see that the components of the optical force can be presented by the term representing the force acting upon an isolated nanosphere located at the position of particle 2 (the scattering force) and the terms including Green's function, $G_{ij}$, describing the coupling force acting between the nanospheres, which is mediated by the light scattering (the optical binding force).

Consider the far field approximation when the distance between particles exceeds the wavelength of the electromagnetic field $(R \gg \lambda_p)$. The far-field design is important in order to prevent re-absorption of photons spontaneously radiated by the RE ions excited by the incident electromagnetic field. As one can see in relation (1c) terms $1/(k_0 R)$ dominate in the far-field approximation. Without loss of generality, one can assume that the incident plane linearly polarized electromagnetic wave has the following components of the electric field: $E_x = 0$, $E_y = E_0 exp(ik_0 z)$, $E_z = 0$ (Fig. 1). Following [27] and substituting these components of the electric field into relation (3) one can find the components of the optical force acting the nanospheres of the dimer in the far filed.

For the dimer placed normally to the light propagation (Fig. 1a) the force applied to nanoparticle 2 along the y-axis looks like

$$F_{y,\perp}^{(2)} = 4\pi\varepsilon_0 |\alpha|^2 |E_0|^2 \frac{k_0^2}{R^2} cos(k_0 R). \tag{4}$$

This binding force oscillates following $cos(k_0 R)$. The stable configuration can be reached for $cos(k_0 R) = 1$ or $R = \lambda_p M$, where $M$ is a natural number. The force applied to nanoparticle 2 presented in Fig. 1a along the z-axis has the form

$$F_{z,\perp}^{(2)} = 2\pi\varepsilon_0 k_0 \alpha" |E_0|^2. \tag{5}$$

For the dimer placed along the direction of the light propagation (Fig. 1b) the force applied to particle 1 and particle 2 along the z-axis, respectively, have the form

$$F_{z,\parallel}^{(1)} = 2\pi\varepsilon_0 k_0 |E_0|^2 \alpha" + 4\pi\varepsilon_0 \frac{k_0^3}{R} |E_0|^2 \left[(\alpha')^2 sin(2k_0 R) + \alpha'\alpha" cos(2k_0 R)\right], \tag{6}$$

$$F_{z,\parallel}^{(2)} = 4\pi\varepsilon_0 \left(\frac{1}{2} k_0 |E_0|^2 \alpha" + \frac{k_0^3}{R} |E_0|^2 \alpha'\alpha"\right). \tag{7}$$

The binding force between the two particles presented in Fig.1b can be estimated as

$$F_{z,\parallel}^{(2-1)} = F_{z,\parallel}^{(2)} - F_{z,\parallel}^{(1)}. \tag{8}$$

RE ions doped in a dielectric host material change the electric permittivity of the material if the incident electromagnetic wave propagates at the frequency (wavelength) closely-placed to the ion resonance frequencies. Indeed, the electric permittivity of RE-dopes dielectric can be presented by the relation [29]

$$\varepsilon_p = \varepsilon_0(1 + \mu_0 + \mu_1 + \cdots), \tag{9}$$

where $\varepsilon = \varepsilon_0(1 + \mu_0) = \varepsilon_0 n_{YAG}^2$ is the electric permittivity of undoped (pure) host material, $n_{YAG}$ is the refractive index of the pure host. In our scheme $n_{YAG}$ is the refractive index of undoped YAG. An expression for susceptibility $\mu_1$ has the form

$$\mu_1 = \frac{\omega_p^2}{\omega_0^2 - \omega^2 + i\sigma\omega}, \quad \text{where} \quad \omega_p^2 = \frac{e^2 N_1}{\varepsilon_0 m}. \tag{10}$$

Here $\omega_p$ is plasma angular frequency, $m$ and $e$ are the effective mass and charge of an electron, respectively. $N_1$ is the concentration of RE ions excited by the incident elrctromagnetic field. $\omega_0$ is the resonance angular frequency of Yb$^{3+}$ ions. $\sigma$ is the damping coefficient near $\omega_0$. For Yb$^{3+}$YAG the damping coefficient is $\sigma = 6.26 \times 10^{12} Hz$ at wavelength $968\ nm$ and $\sigma = 2.34 \times 10^{13} Hz$ at wavelength $1030\ nm$.

Yb$^{3+}$ ions can be considered as a two-level system. The number of the excited ions can be estimated using the rate equations [30]

$$\frac{dN_1}{dt} = \frac{I_p}{h\nu_p}[N_0\sigma_a(\nu_p) - N_1\sigma_e(\nu_p)] - \frac{N_1}{\tau_r} - \frac{N_1}{\tau_{nr}}$$
$$N_T = N_0 + N_1 \tag{11}$$

where $N_T$ is the RE ion density in the sample, $N_0$ and $N_1$ are the population dencity in the ground and exited manifolds, respectively. $I_p$ is the intensity of the incident electromagnetic wave propagating at the frequency $\nu_p = c/\lambda_p$. $\sigma_{a,e}(\nu_p)$ are the absorption (*a*) and emission (*e*) cross sections at the frequency $\nu_p$. $\tau_r$ and $\tau_{nr}$ are the radiative and nonradiative lifetime of the excited level, respectively. Spontaneous emission can be characterized by the mean fluorescence frequency

$$\nu_F = \frac{\int \Phi(\nu)\nu d\nu}{\int \Phi(\nu)d\nu}, \tag{12}$$

where $\Phi(\nu)$ is the fluorescence flux density, which can be obtained experimentally, and $\nu_F = c/\lambda_F$. Here $\lambda_F$ is the mean fluorescence wavelength. It is necessary to stress that if the frequency of the incident electromagnetic field is smaller than the mean fluorescence frequency $(\nu_p < \nu_F)$ and the quantum efficiency is very high $(> 99.8\%)$ the sample undergoes cooling otherwise it undergoes heating [30, 31]. Cooling can be reached only in high quality low phonon host materials [32]. In the simplest case of the ideal sample the heating or cooling energy density generated in the sample by the incident electromagnetic wave can be estimated as [32]

$$E = N_1 hc(\nu_p - \nu_F). \tag{13}$$

## 3. Results and discussion

In this part of the paper, we'll see how RE ions doped into a dimer influence the optical forces acting the nanospheres of the dimer. As noted above, I consider the dimer composed of two identical Yb$^{3+}$:YAG nanospheres. The radius of the nanospheres in my simulations is $a = 110\ nm$. The energy levels of Yb$^{3+}$ ions undergo Stark splitting in the host material (Fig. 2).

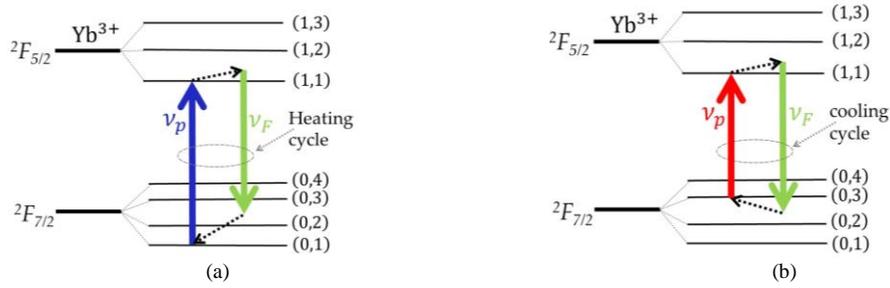

Fig. 2. Energy levels of Yb$^{3+}$ ions doped in a solid. Solid arrows illustrate the excitation and decay processes for Stokes (a) and anti-Stokes (b) cycles. Dotted arrows illustrate phonon generation and phonon absorption. $\nu_p$ is the frequency of the incident light. $\nu_F$ is the mean fluorescence frequency.

Level $^2F_{5/2}$ splits into three sub-levels, level $^2F_{7/2}$ splits into four sub-levels. The energies of sub-levels almost do not change with size of the samples even at the nanosized scale. For nanoparticles the energies of the sub-levels are almost identical to the energies of the sub-levels of bulk Yb$^{3+}$:YAG presented in Ref. [33]. YAG is a low-phonon material with maximum phonon energy $860\ cm^{-1}$. High quantum efficiency($> 99.8\%$) can be achieved in high quality Yb$^{3+}$YAG crystals.

Consider the incident electromagnetic field at two wavelengths: $\lambda_p = 968\ nm$ corresponds to the transition between (0,1) and (1,1) sub-levels, $\lambda_p = 1030\ nm$ corresponds to the transition between (0,3) and (1,1) sub-levels. It is important to emphasize that the incident wavelength $968\ nm$ is less than the mean fluorescence wavelength $1002\ nm$ obtained using relation (12), consequently, the incident light at $\lambda_p = 969\ nm$ is a source of the Stokes cycle accompanied by phonon generation (Fig. 2a). The sample undergoes heating, which is undesirable in many applications. The incident light at $\lambda_p = 1030\ nm$ is a source of the anti-Stokes cycle accompanied by phonon absorption (Fig. 2b). The sample undergoes cooling. In all simulations the steady state system ($d/dt = 0$) has been considered.

*3.1 Forces applied to the dimer in the perpendicular configuration*

Consider the case of the perpendicular configuration when the incident electromagnetic wave propagates normally to the dimer axis (Fig. 1a). As mentioned above, the system is stable if $cos(k_0 R) = 1$ or $R = \lambda_p M$, where $M$ is a natural number. For the wavelengths $\lambda_p = 968\ nm$ and $\lambda_p = 1030\ nm$ the system is stable if $R = M \times 968\ nm$ and $R = M \times 1030\ nm$, respectively. Taking into account the symmetry of the system and using relation (4) one can estimate the force applied to nanoparticle 2 (Fig. 1a) along the y-axis at the wavelengths $\lambda_p =$ 968 nm and $\lambda_p = 1030\ nm$. These forces are presented in Fig. 3. Fig. 3 also illustrates the forces acting on nanoparticle 2 of the dimer made of undoped YAG. These forces were estimated using relation (4).

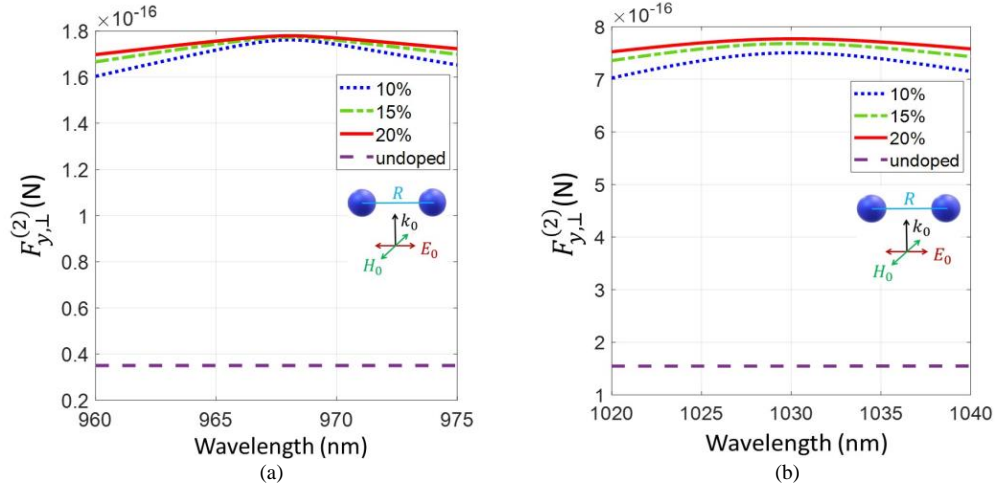

Fig. 3. Force, $F_{y,\perp}^{(2)}$, applied to nanosphere 2 of the dimer made of 10%, 15%, 20% $Yb^{3+}$:YAG, and undoped YAG as a function of the wavelength. The incident light propagates at the wavelength 968 nm (a) and 1030 nm (b) with the intensity $I_p = 1.5 \times 10^9\ W/m^2$.

In Fig. 3 the intensity of the electromagnetic field applied to the dimer is $I_p = 1.5 \times 10^9 (W/m^2)$ for each of the considered wavelengths. The distances between nanospheres are $R = 6776\ nm$ at $\lambda_p = 968\ nm$ and $R = 3090\ nm$ at $\lambda_p = 1030\ nm$. They correspond to the stable systems. As one can see in Fig. 3, the optical forces applied to the $Yb^{3+}$ doped dimer significantly exceed the optical forces applied to the identical undoped dimer made of YAG. This effect is caused by the change in the polarizability of the nanospheres of the dimer represented by relation (2). It is worth to remind that the incident field at wavelength $\lambda_p = 968\ nm$ heats the dimer, while the incident field at $\lambda_p = 1030\ nm$ cools it as a result of the Stokes and anti-Stokes cycles, respectively (Fig. 2). For example, if the intensity of the incident field is $I_p = 1.5 \times 10^9\ W/m^2$ and the $Yb^{3+}$ ion concentration is 15% the heating energy density and cooling energy density generated in an ideal sample are approximately $E = 9.6 \times 10^{-18}\ J/m^3$ and $E = 4.5 \times 10^{-18}\ J/m^3$, respectively. These values have been estimated using relation (13). $Yb^{3+}$ ions doped in the YAG dimer also increase the scattering force applied to the dimer along the direction of the propagation of the incident electromagnetic wave (Fig. 4). The scattering force has been estimated using relation (5). As one can see in Fig. 3 and Fig. 4 the concentration of the $Yb^{3+}$ ions slightly influence the value of optical force applied to $Yb^{3+}$:YAG dimer. All identical optical forces applied to the $Yb^{3+}$:YAG dimer at $\lambda_p = 968\ nm$ and $\lambda_p = 1030\ nm$ are comparable with each other and considerably exceed the identical optical forces acting on the undoped YAG dimer at the corresponding wavelengths.

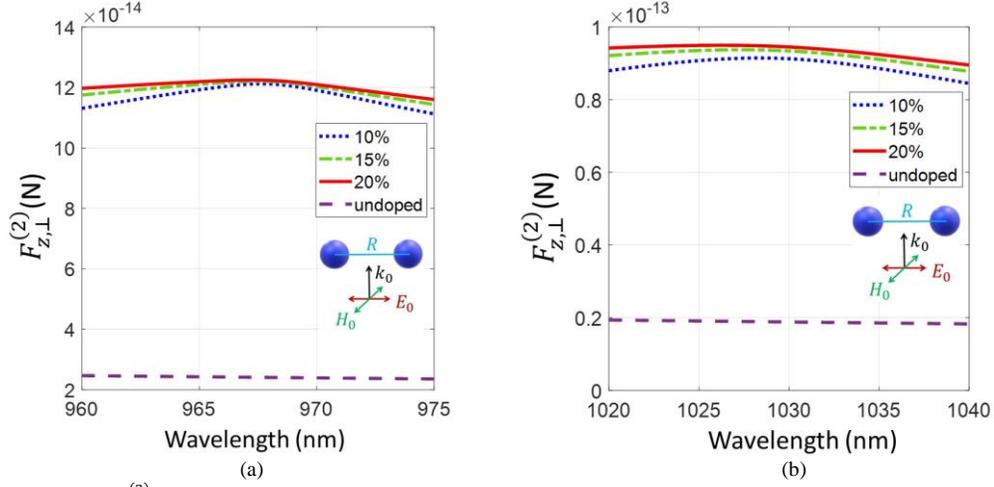

Fig. 4. Force, $F_{z,\perp}^{(2)}$, applied to nanosphere 2 of the dimer made of 10%, 15%, 20% $Yb^{3+}$:YAG, and undoped YAG as a function of the wavelength. The incident light propagates at the wavelength 968 nm (a) and 1030 nm (b) with the intensity $I_p = 1.5 \times 10^9 \, W/m^2$.

The optical force applied to the dimer can be controlled with the intensity of the incident light. Fig. 5 illustrates force $F_{z,\perp}^{(2)}$ applied to the dimer for three different intensities of the incident light $I_p = 1.5 \times 10^9 \, W/m^2$, $I_p = 2.25 \times 10^9 \, W/m^2$, and $I_p = 3 \times 10^9 \, W/m^2$. In Fig. 5 the $Yb^{3+}$ concentration is 15%. As one can see in relation (11) and relation (10) the light intensity controls the number of the excited $Yb^{3+}$ ions, $N_1$. The number of the excited ions influence susceptibility of the nanospheres of the dimer, $\mu_1$.

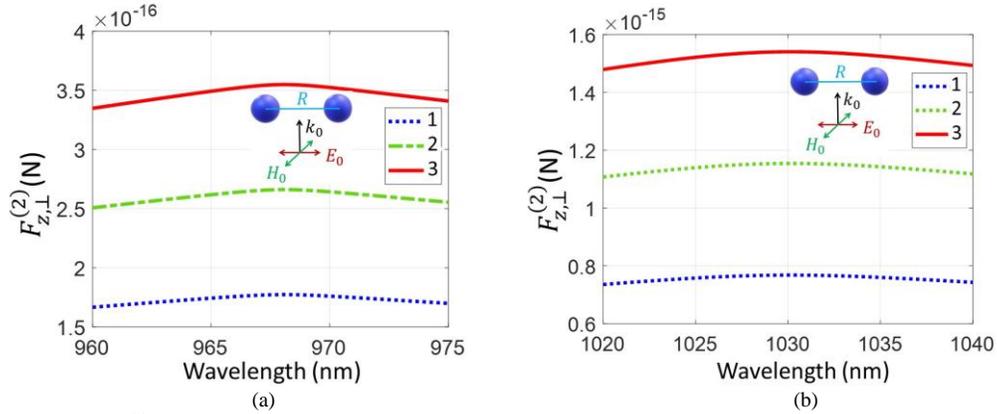

Fig. 5. Force, $F_{z,\perp}^{(2)}$, applied to nanosphere 2 of the $Yb^{3+}$:YAG dimer as a function of the wavelength at light intensities $I_p = 1.5 \times 10^9 \, W/m^2$, (1), $I_p = 2.25 \times 10^9 \, W/m^2$, (2), and $I_p = 3 \times 10^9 \, W/m^2$, (3). The incident light propagates at the wavelength 968 nm (a) and 1030 nm (b).

*3.2 Forces applied to the dimer in the longitudinal configuration*

Consider the case of the longitudinal configuration when a dimer placed along the direction of the propagation of the incident light (Fig. 1b). This system is stable if $cos(2k_0 R) = 1$ or $R = M \lambda_p/2$, where $M$ is a natural number. Consider the stable dimers with $R = 6776 \, nm$ at $\lambda_p = 968 \, nm$ and $R = 3090 \, nm$ at $\lambda_p = 1030 \, nm$. The optical forces applied to particle 1, $F_{z,\parallel}^{(1)}$, and particle 2, $F_{z,\parallel}^{(2)}$, along the direction of the propagation of the incident light (z-axis) can be estimated using relations (6) and (7), respectively. The values of these forces are comparable. The optical forces applied to particle 1, $F_{z,\parallel}^{(1)}$, is presented in Fig. 6. As one can see in Fig. 6, $F_{z,\parallel}^{(1)}$ slightly depends on the concentration of Yb$^{3+}$ ions. The optical force applied to the RE doped dimer made of Yb$^{3+}$:YAG essentially exceeds the identical optical force applied to the undoped dimer made of YAG (Fig. 6). The intensity of the incident electromagnetic field at both wavelengths is $I_p = 1.5 \times 10^9 \, W/m^2$.

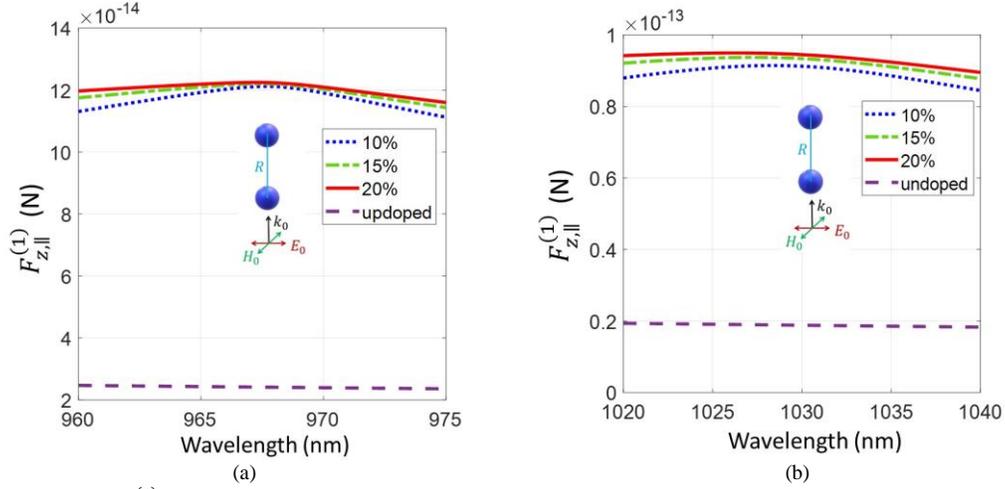

Fig. 6. Force, $F_{z,\parallel}^{(1)}$, applied to nanosphere 1 of the dimer made of 10%, 15%, 20% Yb$^{3+}$:YAG, and undoped YAG as a function of the wavelength. The incident light propagates at the wavelength 968 nm (a) and 1030 nm (b) with the intensity $I_p = 1.5 \times 10^9 \, W/m^2$.

The force binding two nanospheres of the dimer has been estimated using relation (8). It is presented in Fig. 7. The negative value of the binding force means that it is opposite in direction to the z-axis.

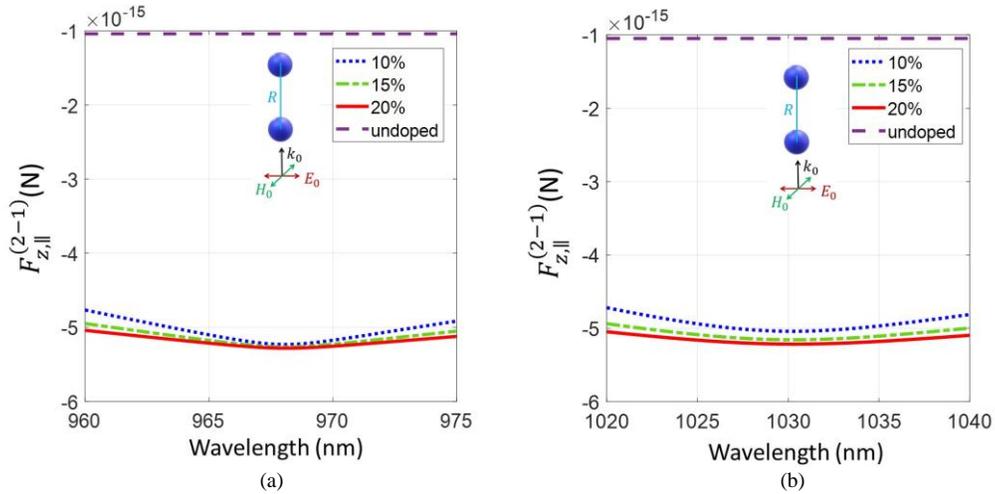

Fig. 7. Binding force, $F_{z,\parallel}^{(2-1)}$, acting in 10%, 15%, 20% Yb$^{3+}$:YAG, and undoped YAG dimer as a function of the wavelength. The incident light propagates at the wavelength 968 nm (a) and 1030 nm (b), with the intensity $I_p = 1.5 \times 10^9 \, W/m^2$.

In common with the case of the perpendicular configuration, the intensity of the incident light can be used to control the optical forces in the system. As an example, Fig. 8 illustrates the influence of the intensity of the incident light on the binding force. In Fig. 8 the binding force $F_{z,\parallel}^{(2-1)}$ has been simulated for three different light intensities $I_p = 1.5 \times 10^9 \, W/m^2$, $I_p = 2.25 \times 10^9 \, W/m^2$, and $I_p = 3 \times 10^9 \, W/m^2$. The concentration of $Yb^{3+}$ ions is 15%.

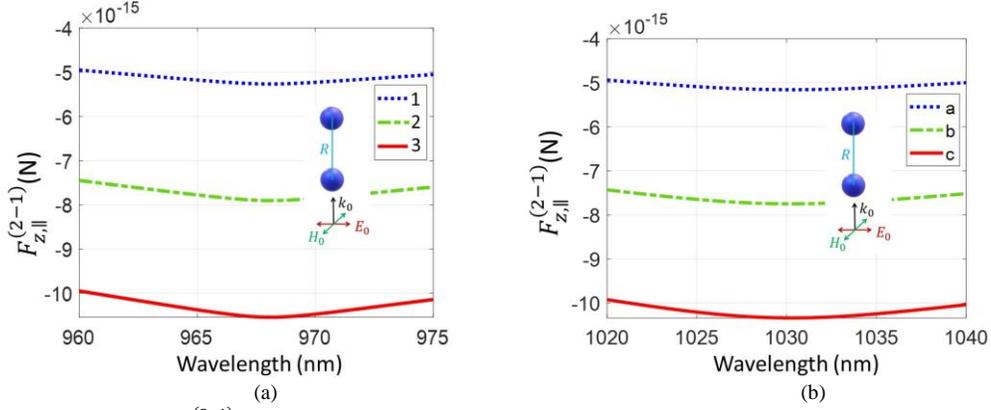

Fig. 8. Binding force, $F_{z,\parallel}^{(2-1)}$, applied to the $Yb^{3+}$:YAG dimer as a function of the wavelength at light intensities $I_p = 1.5 \times 10^9 \, W/m^2$, (1), $I_p = 2.25 \times 10^9 \, W/m^2$, (2), and $I_p = 3 \times 10^9 \, W/m^2$, (3), respectively. The incident light propagates at the wavelength 968 nm (a) and 1030 nm (b).

## 4. Conclusions

In this work the optical forces applied to an $Yb^{3+}$:YAG dimer composed of two identical nanospheres with radii $110 \, nm$ have been investigated. The optical forces applied to the RE doped dimer made of $Yb^{3+}$:YAG and the optical forces applied to the identical undoped dimer made of YAG have been compared. Two schemes have been considered. In one of the schemes the incident plane, linearly polarized electromagnetic wave propagates normally to the dimer axis; in the second one the incident plane, linearly polarized wave propagates along the dimer axis. The incident light propagating at wavelength $\lambda_p = 968 \, nm$ and $\lambda_p = 1030 \, nm$ has been considered in both schemes. These wavelengths correspond to the electron transitions of the $Yb^{3+}$ ions. The distance between the nanospheres, $R$, was chosen to get a stable configuration of the system at the corresponding incident wavelength. This distance is $R = 6776 \, nm$ at the incident wavelength $\lambda_p = 968 \, nm$ and $R = 3090 \, nm$ at the wavelength $\lambda_p = 1030 \, nm$. The incident light propagating at the wavelength $\lambda_p = 968 \, nm$ generates heat in the $Yb^{3+}$:YAG dimer, while the incident light propagating at $\lambda_p = 1030 \, nm$ can cool the dimer made of the high purity $Yb^{3+}$:YAG. A bulk $Yb^{3+}$:YAG sample was laser cooled in air starting from room temperature [34]. It has been shown that the values of the optical forces acting on the nanospheres of the dimer at the incident wavelengths $\lambda_p = 968 \, nm$ and $\lambda_p = 1030 \, nm$ are comparable with each other and exceed the identical optical forces applied to the identical undoped dimer at the same wavelengths. By this means RE ions can not only be use to increase the optical forces acting on the dimer but, at the properly chosen wavelength of the incident light, RE ions can cool the dimer. This is a very promising effect in a number of applications, where heat generated in the dimer is undesirable. This approach is not limited to $Yb^{3+}$:YAG material, since optical cooling has been observed in many low phonon materials doped with $Yb^{3+}$ ions, among them ZBLANP, YLF, YSO, ABCYS [32, 35]. The results obtained in the paper can be interesting to experimentalists working in the area of simultaneous light manipulation of nanoparticles.